\documentclass[a4,
 reprint,%
 amssymb, amsmath,%
 aip,
 unsortedaddress,
]{revtex4-1}

\usepackage{amsmath,amsfonts,amssymb}
\usepackage{graphicx}
\usepackage[colorlinks=true, allcolors=blue]{hyperref}

\begin{document}
\title{Prospects of designing gold-nanoparticles-based soft terahertz radiation sources
and terahertz-to-infrared converters for concealed object detection technology}

\author{K. A. Moldosanov}
\email{altair1964@yandex.ru}
\affiliation{Kyrgyz-Russian Slavic University, 44 Kiyevskaya St., Bishkek 720000, Kyrgyzstan}
\author{A.~V.~Postnikov}
\affiliation{LCP-A2MC, University of Lorraine, 1 Bd Arago, F-57078 Metz, France}
\author{V.~M.~Lelevkin and N.~J.~Kairyev}
\affiliation{Kyrgyz-Russian Slavic University, 44 Kiyevskaya St., Bishkek 720000, Kyrgyzstan}

\begin{abstract}
The two-phonon scheme of generation of terahertz (THz) photons by gold nanobars (GNBs) is considered.
It is shown that in GNBs, by choosing their sizes, it is possible to provide conditions for converting
the energy of longitudinal phonons with THz frequencies into the energy of THz photons. 
The prospects of designing GNBs-based soft THz radiation sources (frequencies: 0.14; 0.24; 0.41 and 0.70 THz)
with a large flow cross-section (diameter $\sim\,$40~cm) intended for detection of hidden objects
under clothing to ensure security in public places (airports, railway stations, stadiums, etc.)
are assessed. The choice of the above frequencies is a compromise between the requirements of
low absorption of THz radiation by water vapor in air, good penetration through the fabric of clothing,
favoring a sufficient resolution of the imaging system, and an abundance of corresponding 
longitudinal phonons, capable of exciting Fermi electrons in GNBs. Estimates of the characteristics 
of the terahertz-to-infrared converter based on gold nanospheres (GNSs), which could work 
in tandem with these sources of THz radiation -- as a means of visualization of hidden objects -- are also given. 
\end{abstract}


\maketitle

\section{Introduction}
\label{sec:intro}  
\begin{figure}[!t]
\begin{center}
\includegraphics[width=6.0cm]{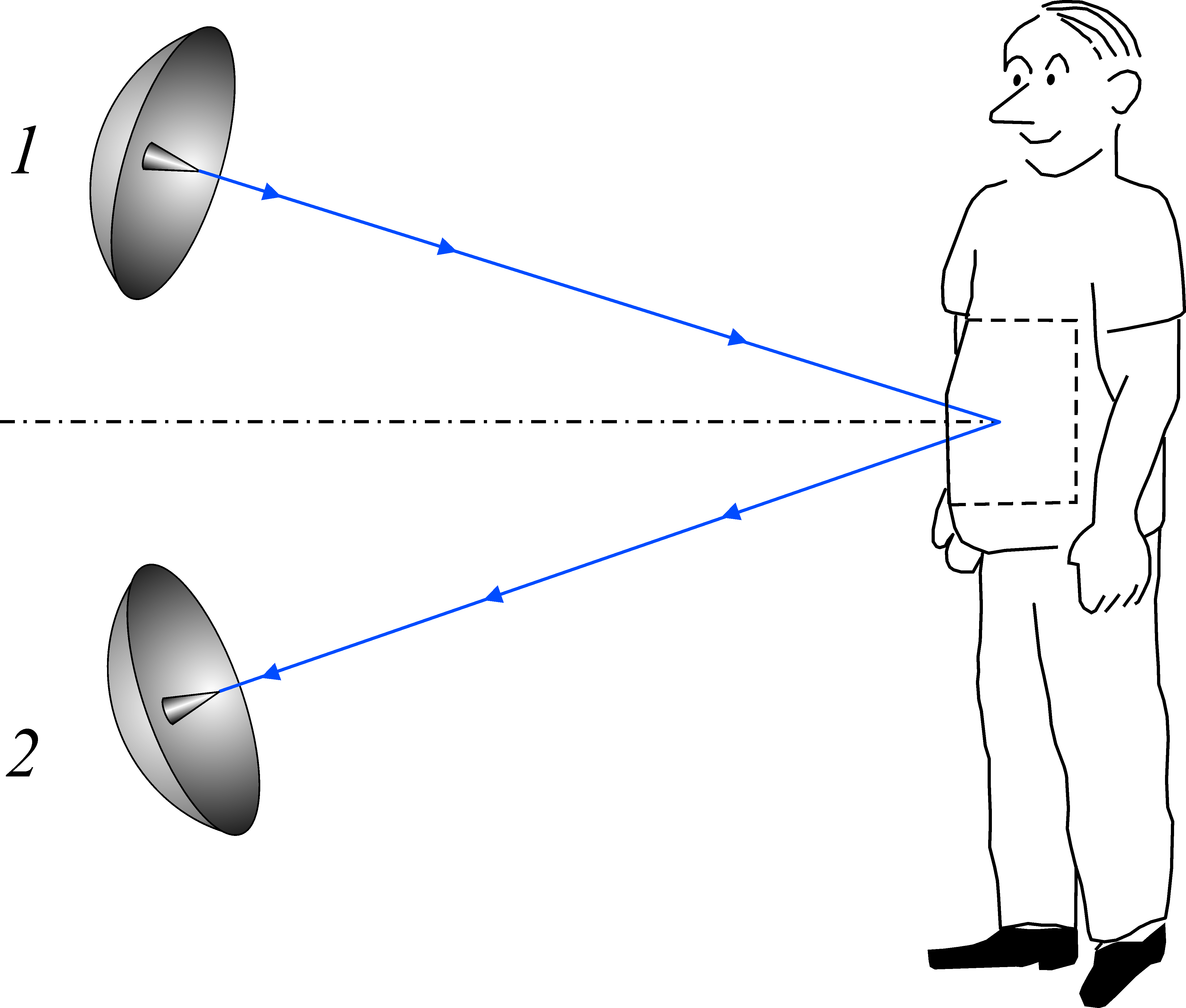}
\end{center}
\caption[example]{\label{fig:01} 
Typical geometry of the remote threat detection. 1: THz radiation source, 2: THz radiation detector.}
\end{figure}

From the very beginning of practical exploring the terahertz (THz) range, attempts have been made
to apply it for remote detection of hidden threats on the human body -- see, e.g., 
Ref.~\citenum{APL85-519,ProcSPIE5354-Tribe,ProcSPIE6212-62120E,ProcSPIE6402-64020D,IEEEtransAntePropa55-2944,%
ProcSPIE6948-69480M,IEEEconfHST09-440,Moeller_report233347,IETcolloqMWTET-Robertson,ProcSPIE7671-76710Y,
IEEEtransTeraSciTech1-282,AIPR05-Linden,Daniels_EMdetection,TST9-19}
and references therein. Fig.~\ref{fig:01} illustrates the geometry of active remote weapon detection
(gun, bomb, knife, etc.) hidden under the clothes. In this method of detecting threats on the subject's body,
the latter is irradiated by an external source of THz radiation, and radiation reflected from hidden objects
is recorded. Previous works allowed to identify the main problems complicating 
a successful application of THz radiation for the remote detection of hidden objects,
namely, an attenuation of THz radiation in fabrics of clothes, and its absorption by water vapor in the air. 
It was found that the highest transmittance of clothing tissue is in the low-frequency part 
($\sim\,0.1-0.5$~THz) of the THz range. In this range, the attenuation of THz radiation in air water vapor is relatively low, but monotonically increases with increasing frequency and is saturated
with damping peaks due to the absorption by oscillating hydrogen atoms in the water molecule. The latter limits the detection range; however, in most hidden object detection systems, the detection range is anyway relatively short, from 1 to 20 m.\cite{Daniels_EMdetection}
As for the application for short-range operation ($<\,$4~m), such as airport security check,
the lower operating frequency ($\sim\,0.1$~THz) can still provide diffraction-limited image resolution
of $\sim\,$2~cm, which should suffice for many threat scenarios.

\begin{table*}[!t]
\caption{\label{tab:01}
Some numerical parameters for photons at four frequencies selected for our analysis
}
\begin{center}
\begin{tabular}{|r@{.}l|r@{.}l|r@{.}l|r@{.}l|r@{.}l|r@{.}l|}
\hline
 \multicolumn{2}{|c}{Frequency} & 
 \multicolumn{2}{|c}{Wavelength} & 
 \multicolumn{2}{|c}{Photon energy} & 
 \multicolumn{2}{|c}{Photon momentum} &
 \multicolumn{2}{|c}{Objective resolution$^{\ast}$} &
 \multicolumn{2}{|c|}{Penetration} \\*[-2pt]
 \multicolumn{2}{|c}{$\nu$ (THz)} & 
 \multicolumn{2}{|c}{(mm)} &
 \multicolumn{2}{|c}{$h{\nu}$ (meV)} &
 \multicolumn{2}{|c}{$p_{\rm ph}$ ($10^{-26}$~g$\,$cm$\,$s$^{-1}$)} &
 \multicolumn{2}{|c}{${\Delta}x\,{\approx}\,{\lambda}d/A$ (cm)} &
 \multicolumn{2}{|c|}{depth$^{\dagger}$ ($\mu$m)} \\
\hline
 \hspace*{5mm}  0&14  & 
 \hspace*{6mm}  2&1   & 
 \hspace*{8mm}  0&58  & 
 \hspace*{12mm} 3&1   & 
 \hspace*{12mm} 4&2   &
 \hspace*{6mm}  0&20  \\
 0&24  & 1&25 & 0&99 & 5&3  &           2&5 & 0&15 \\
 0&41  & 0&73 & 1&70 & 9&1  & ${\sim}\,$1&5 & 0&12 \\
 0&70  & 0&43 & 2&90 & 15&5 & ${\sim}\,$0&9 & 0&09 \\
\hline
\end{tabular} \\
$^{\ast}$for $d{\,\sim\,}10$~m, $A{\,\sim\,}0.5$~m; \rule[0mm]{0mm}{4mm}
\hspace*{6mm}
$^{\dagger}$skin layer thickness in bulk Au
\end{center}
\end{table*}

The development of active methods for detecting hidden objects revealed a shortage of sources
of soft THz radiation with a wide radiation flux. Typically, the THz radiation source would be required
to create a spot on the human body with a diameter of $\sim\,$40~cm. Accordingly, the imaging system
aperture $A$ must be at least this size. As a rule, THz radars with an aperture of $\sim\,50-100$~cm
are used,\cite{ProcSPIE6948-69480M,IEEEconfHST09-440,Moeller_report233347,IETcolloqMWTET-Robertson,%
ProcSPIE7671-76710Y,IEEEtransTeraSciTech1-282} 
which is a compromise between the resolution ${\Delta}x$ of the imaging system,
the range of observation of the subject $d$ and the THz radiation wavelength $\lambda$, in accordance
with the formula for a diffraction-limited resolution: ${\Delta}\,x\,\approx\,\lambda\,d/A$. Naturally,
the wavelength $\lambda$ is chosen so that it is not at the same time at the peak of attenuation 
in the atmospheric air and belongs to the region of acceptable transmission of clothing fabrics, that is,
the compromise wavelengths correspond to frequencies ${\sim\,}0.1-0.5$~THz. The resolution accessible
is within the range ${\Delta\,x\,\sim\,}0.5-2$~cm.

\begin{figure}[!b]
\begin{center}
\includegraphics[width=0.48\textwidth]{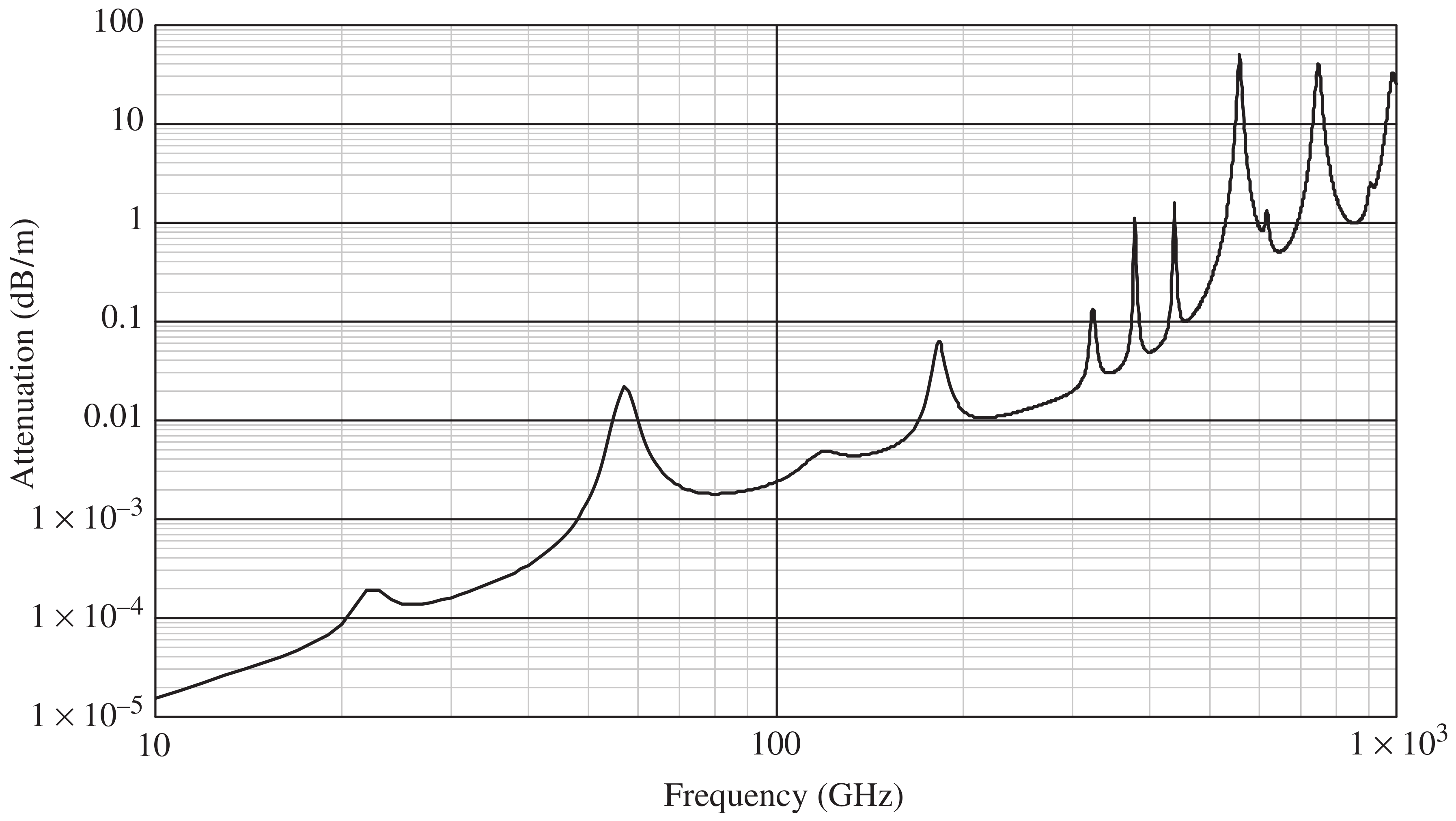}
\end{center}
\caption[example]{\label{fig:02} 
Electromagnetic waves attenuation by an air layer of 1~m thickness, in the frequency range 0.01 through 1 THz,
reproduced from Figure~2.15 of Ref.~\citenum{Daniels_EMdetection}.
The frequencies chosen for our analysis are 0.14; 0.24; 0.41 and 0.70 THz.}
\end{figure} 

In the present paper we suggest the novel soft terahertz radiation sources based on gold nanobars (GNBs)
being irradiated with microwaves. This approach assesses the feasibility of the idea to convert
the energy of longitudinal phonons into electromagnetic energy of THz photons. The approach
seems to be reinforced by recent reports\cite{Nature561-216,Nature567-E12,ACSphotonics5-3082} 
on an enhancement beyond the blackbody limit of radiative heat transfer in nanometric-scale objects,
as well as by our own estimates of the surface
power density of spontaneous THz radiation by gold nanoparticles (see Appendix A).

To enhance the power of THz radiation, GNBs provide the conditions (see Appendix B) to increase
the number of longitudinal phonons by heating GNBs with microwave radiation. This approach would enable
a wide beam of THz radiation to be issued from a large number of GNBs distributed on a substrate
or within a matrix of sufficiently large size.

Taking into account the data of 
Ref.~\citenum{IEEEtransTeraSciTech1-282,AIPR05-Linden,Daniels_EMdetection} 
on the attenuation of THz radiation in the air, we have chosen
the following frequencies to detect hidden objects: 0.14; 0.24; 0.41 and 0.70 THz 
(the corresponding wavelengths, energies and momenta of THz photons are given in Table~\ref{tab:01}).
These frequencies do not fall into the attenuation peaks of THz radiation in air -- 
see Fig.~\ref{fig:02}, borrowed from Ref.~\citenum{Daniels_EMdetection}.
In addition, at these frequencies
the undesirable tissue absorption is sufficiently low, whereas the desired transmission, on the contrary,
is high.\cite{IEEEtransTeraSciTech1-282,JNanophoton6-061716}
Table~\ref{tab:01} includes also the objective's resolution ${\Delta}x$, achievable at
chosen frequencies, with the distance to the subject $d{\,\sim\,}10$~m and the aperture $A{\,\sim\,}0.5$~m.

The paper is organized as follows. Sec.~\ref{sec:02} explains generation of THz radiation by gold nanobars,
Sec.~\ref{sec:03} suggests the practical design of the THz source. In Sec.~\ref{sec:04}, the parameters
of the GNSs-based THz-to-IR converter, which could be used to visualize hidden objects
at the security checkpoints, are briefly discussed. The paper is concluded by Sec.~\ref{sec:conclu}
and contains two Appendices, A and B.

\section{Suggested design of soft terahertz radiation sources
with gold nanobars as active elements}
\label{sec:02}
In the proposed approach, in a GNB, the Fermi electron is excited by absorbing a longitudinal phonon, 
and relaxes by releasing a softer longitudinal phonon. The energy difference is brought away by the electron 
and emitted as a soft THz photon as the electron scatters at the GNB boundary. The physical picture to be considered
is somewhat similar to that discussed earlier 
in Ref.~\citenum{BeilJNano7-983,Patent-RU2622093,Patent-RU2650343}, 
with the only difference that in these works the phonon energy was completely converted 
into the energy of THz phonon. As emphasized in the works cited, the role of gold nanoobjects
is to provide free electrons and longitudinal phonons, their respective energy levels
being appreciably discretized due to spatial confinement. The microwave radiation
serves as a source of energy for maintaining the ``phonon bath'' that heats the nanoobjects.

\begin{figure*}[!t]
\begin{center}
\includegraphics[width=0.8\textwidth]{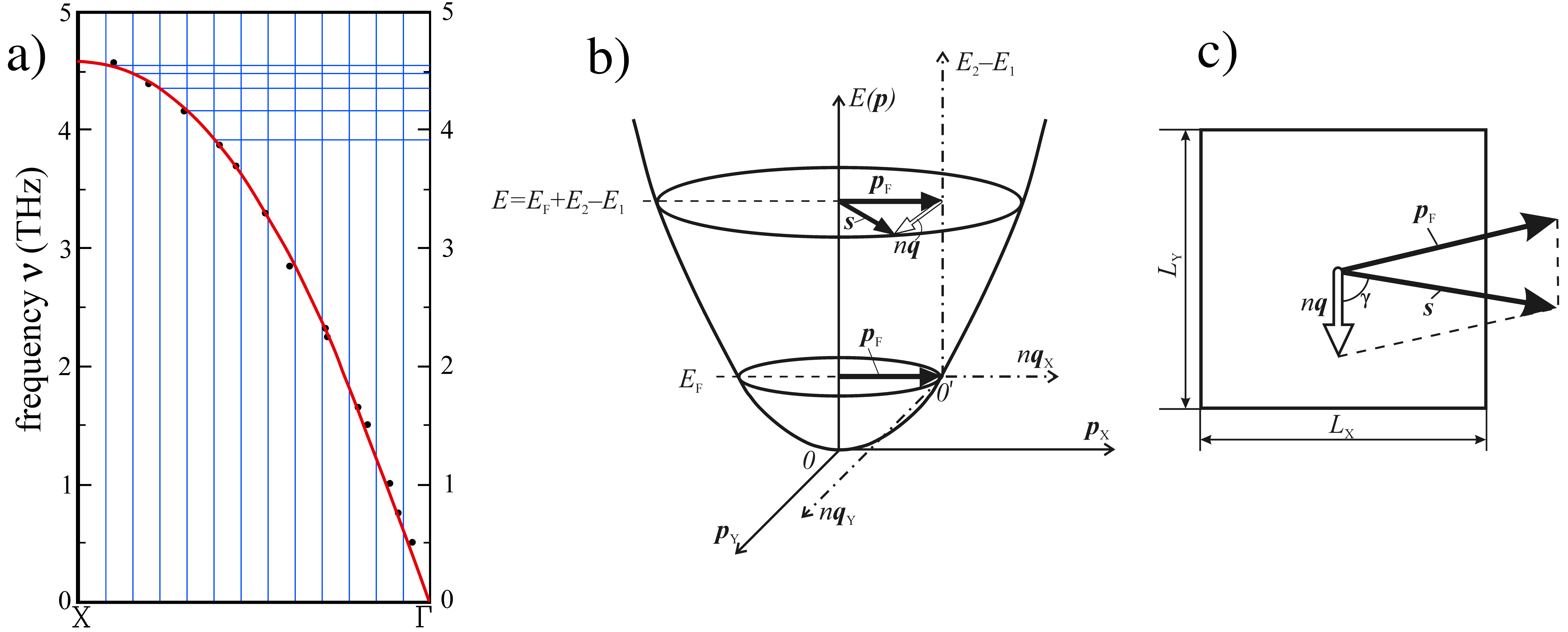}
\end{center}
\caption[example]{\label{fig:03} 
Energy and momentum relations concerning absorption and subsequent emission of a longitudinal
phonon by a Fermi electron. 
(a) Phonon dispersion in bulk gold along $X\,-\,{\Gamma}$ (red line), discretized
as a sequence of spatial confinement (assuming $N_{x,y}=13$) in a ``thin'' GNB (vertical blue lines
for wavevectors, horizontal blue lines for frequencies). 
(b) Energy of a Fermi electron as function of $(x,y)$ momenta prior to and after the absorption / 
emission of longitudinal phonons with energies $E_2$ / $E_1$. $\mathbf{p}_{\rm F}$ is the Fermi momentum,
$n\mathbf{q}$ the difference of phonon momenta, $\mathbf{s}$ the momentum of the electron
after the absorption / emission of phonons. 
(c) The orientation of the three momenta $\mathbf{p}_{\rm F}$, $n\mathbf{q}$, $\mathbf{s}$,
making clear the meaning of the angle $\gamma$.
See text for detail.
}
\end{figure*}

\subsection{General considerations}
To streamline the argumentation, it seems useful to consider a simplified picture 
in which the wavevector of the primary (exciting) phonon, $\mathbf{q}_2$,
and the phonon released in the course of the electron relaxation, $\mathbf{q}_1$,
are collinear and directed along a short --
just several lattice parameters of gold $a_{\rm Au}$ -- GNB dimension,
say $L_X=N_x\,a_{\rm Au}$ or $L_Y=N_y\,a_{\rm Au}$ (see Fig.~\ref{fig:03}),
so that the spatial confinement results in discretisation of $(x,y)$ phonon momenta.
Now, we assume that the phonon dispersion of bulk gold still approximately holds
at the nanoscale (this seems plausible since the vibration spectra of bulk and
nanoparticle gold are very close -- see Fig.~2 of Ref.~\citenum{BeilJNano7-983}
and the original papers\cite{PRB8-3493,PRB87-014301,Nanoscale6-9157}), and we want to deduce some
conclusions from matching the conditions of electron and phonons momenta and energy
conservation. Such an analysis has been done in Ref.~\citenum{Nanotechnology29-285704},
albeit with compact GNPs (not elongated GNBs) in mind. The discussion around Fig.~2 
in Ref.~\citenum{Nanotechnology29-285704} 
parametrised the experimental ${\Gamma}-X$ dispersion relation for bulk gold
and elaborated on the matching conditions between the energy of absorbed/released phonons, 
$\hbar\omega(\mathbf{q}_1)-\hbar\omega(\mathbf{q}_2)$, and the quantisation step
on the electron excitation.

\begin{table}[!b]
\caption{\label{tab:02}
Energy differences on absorption / emission of longitudinal phonons in gold,
consistently with the dispersion relation as shown in Fig.~\ref{fig:03}. See text for details.}
\begin{center}
\begin{tabular}{|ccc|r@{.}l|r@{.}l|c|}
\hline
\multicolumn{3}{|c|}{\rule[-0pt]{0pt}{12pt}
Wavevectors in units of $\tfrac{(X{\rightarrow}{\Gamma})}{13}$} &
\multicolumn{2}{c|}{} & \multicolumn{2}{c|}{} & \\*[-2pt] 
\hspace*{7mm} & $q_2$ & $q_1$ & \multicolumn{2}{c|}{$E_2\!-\!E_1$~(meV)} & \multicolumn{2}{c|}{$\nu$~(THz)} &
$\gamma$  \\
\hline
\multicolumn{8}{|c|}{\rule[-4pt]{0pt}{14pt} $q_2\!-\!q_1=1.25{\times}10^{-20}$~g$\,$cm$\,$s$^{-1}$} \\
 & 1 & 2 & \hspace*{8mm} 0&59 & \hspace*{3mm} 0&14 & 87$^{\circ}$  \\
 & 2 & 3 &               0&60 &               0&15 & 87$^{\circ}$  \\
 & 3 & 4 &               0&71 &               0&17 & 87$^{\circ}$  \\
 & 4 & 5 &               0&99 &               0&24 & 87$^{\circ}$  \\
\hline
\multicolumn{8}{|c|}{\rule[-4pt]{0pt}{14pt} $q_2\!-\!q_1=2.50{\times}10^{-20}$~g$\,$cm$\,$s$^{-1}$} \\
 & 1 & 3 &               1&19 &               0&29 & 84$^{\circ}$  \\
 & 2 & 4 &               1&31 &               0&32 & 84$^{\circ}$  \\
 & 3 & 5 &               1&70 &               0&41 & 84$^{\circ}$  \\
\hline
\multicolumn{8}{|c|}{\rule[-4pt]{0pt}{14pt} $q_2\!-\!q_1=3.75{\times}10^{-20}$~g$\,$cm$\,$s$^{-1}$} \\
 & 1 & 4 &               1&90 &               0&46 & 82$^{\circ}$  \\
 & 2 & 5 &               2&30 &               0&56 & 82$^{\circ}$  \\
\hline
\multicolumn{8}{|c|}{\rule[-4pt]{0pt}{14pt} $q_2\!-\!q_1=5.00{\times}10^{-20}$~g$\,$cm$\,$s$^{-1}$} \\
 & 1 & 5 &               2&90 &               0&70 & 79$^{\circ}$  \\
\hline
\end{tabular}
\end{center}
\end{table}

In the following, these energy / momentum relations are expressed in a more straightforward way,
assuming $N_x=N_y=13$ for the reasons explained in Appendix B. 
Then $L_X=L_Y=5.3$~nm, and the momenta of longitudinal phonons are quantized
with the step $h/L_X\,\approx\,1.25{\cdot}10^{-20}$~g$\,$cm$\,$s$^{-1}$.
Fig.~\ref{fig:03} summarizes relations between momenta and energies of longitudinal
phonons and of Fermi electrons, in the course of absorption / emission of
phonons. Fig.~\ref{fig:03}a depicts the Au dispersion branch along $X-{\Gamma}$,
divided into $h/L_X$ steps. Different combinations of absorption and emission
phonon wavevectors are shown in Table~\ref{tab:02}, along with the corresponding
energy differences, expressed also in units of frequency. 
Of primary interest are the phonons whose frequencies fall into
the full width at half maximum (FWHM) of the longitudinal phononic density of modes
(shown, e.g., in Fig.~2 of Ref.~\citenum{BeilJNano7-983})
between $\sim\,3.9$~THz (16.2~meV) and $\sim\,4.6$~THz (19.0~meV),
the maximum of the peak in the density of modes being at $\approx\,4.2$~THz (17.4~meV).
The wavevectors of such phonons are close to the Brillouin zone boundary, e.g., 
separated by just few quantisation steps from the $X$ point.
The slope of the frequency/wavevector curve, i.e. the longitudinal speed of sound in this frequency
region which will be used in the following to estimate the energy matching conditions,
is $v^{\ast}_{\rm L}\,\approx\,1{\cdot}10^5$~cm$\,$s$^{-1}$, i.e., considerably reduced 
in comparison with the nominal (zone-center)
speed of sound in bulk gold. The combinations of absorption / emission phonon energies
give rise to the energy gain to be transferred to the THz photon (see discussion below);
the frequency values appearing in Table~\ref{tab:02} cover the range of ``useful''
frequencies from 0.14 to 0.70~THz discussed in Section~\ref{sec:intro}.

Fig.~\ref{fig:03}b depicts the electron energy levels of the GNP, indicating
explicitly a promotion of the Fermi electron (with momentum $\mathbf{p}_{\rm F}$
and energy $E_{\rm F}$), following an absorption and emission of two phonons with wavevectors
close to the Brillouin zone boundary, to an excited state (with momentum $\mathbf{s}$
and energy $E_{\rm F}\!+\!E_2\!-\!E_1$). The relaxation from this excited state will bring about
an emission of a THz photon, as argued below. For the moment, we can make two
remarks, to elaborate on the electron-phonon part.

The first remark is that the energy of the electron excited state $E_{\rm F}\!+\!E_2\!-\!E_1$ must, 
in principle, match one of the discrete levels quantized due to confinement within the GNB, 
in the spirit of the Kubo formula.\cite{JPSJ17-975,JPhysColloq38-C2-69}
The step in the ``ladder'' of electron excitation energies is about
${\Delta}E_{\rm el}\,\approx\,3.38{\cdot}10^{-3}$~meV, as argued in Appendix B.
However, the uncertainty in the phonon energies due to the Heisenberg ratio is about
${\delta}E_{\rm vibr}\,{\sim}\,v^{\ast}_{\rm L}{\hbar}/L_{\rm X}\,\approx\,0.125$~meV,
that makes the above mentioned discreteness of the electron spectrum practically irrelevant,
in the context discussed.

The second remark, illustrated by Fig.~\ref{fig:03}c, is that the electron modifies
the direction of its momentum only weakly, and the Fermi momentum of an electron
participating at the phonon absorption / emission process is roughly at right angle
to the (difference) momentum of the phonons involved. Indeed, 
\begin{eqnarray}
p_{\rm F}^2 &=& s^2+(nq)^2-2s(nq)\cos\gamma\,,\quad\quad \mbox{and} \nonumber \\
\gamma &=& \arccos\frac{s^2+(nq)^2-p_{\rm F}^2}{2s(nq)} \nonumber \\
&=& \arccos\frac{2m(E_2-E_1)+(nq)^2}{2s(nq)}\,.
\end{eqnarray}
The calculated values of $\gamma$ for different combinations of absorbed/emitted phonons
are given in Table~\ref{tab:02}.

\subsection{Channeling the electron excitation energy into emission of soft THz photons}
Following an absorption / emission of ``nearly zone-boundary'' longitudinal phonons
as elaborated above, one can, in principle, imagine the decay of the excited electron state
via promoting an electron across the nanoobject, or via emitting a low-energy phonon,
or via a radiative transition. The last process is of interest for us, as it will be the source
of soft THz radiation. Here we try to present the argumentation why the two other processes 
might turn up to be less likely.

The electron mean free path in gold nanoobjects, expectedly shorter than that in
nearly perfect bulk crystal, seems to be comparable with the GNB cross-section size;
Sec.~9.3.3 of Ref.~\citenum{FANEM2015} discusses this issue, in relation to small gold nanoparticles.
The situation is therefore likely that the electron will arrive at the GNB surface after none, or very few,
scattering events.   
The electron exit from the GNB is precluded by a prohibitively high value of the work function
of gold (4.3~eV\cite{AshMerm_book}), in comparison with the energy values 
(${\simeq}\,1$~meV, see Table~\ref{tab:02}) under discussion.
This energy could have been spent onto an emission of a low-energy phonon. The ``problem'' in this
relation is that this ought to be a nearly zone-center phonon, with the nearly linear dispersion
characterized by the ``nominal'' longitudinal speed of sound in gold, 
$v_{\rm L}=3.23{\cdot}10^5$~cm/s.\cite{IJAPM3-275}
However, the slope of $\omega(q)$ near the center of the Brillouin zone is therefore about 3 times more steep
than that near the phonon energies of ${\simeq}4$~meV, within the FWHM of the peak in the
longitudinal density of vibration modes.
Due to the quantisation of $q$ depicted in Fig.~\ref{fig:03}a, the minimal energy ``quantum''
to excite a low-energy phonon (near $\Gamma$) would be 
${\Delta}E_{\rm vibr}=v_{\rm L}(h/L_X)\,\simeq\,2.52$~meV.
The inspection of $(E_2\!-\!E_1)$ values in Table~\ref{tab:02} shows that such excitation energies 
do not come about before arriving at combinations with $q_1=5$. One can note in this relation
that the issue of the Heisenberg's uncertainty, that was earlier helpful for us to demonstrate
that the exact (anyway difficult to control) quantisation of electron energies is of little
practical importance, won't work the same way for phonons: the quantisation step in the
vibration energies is ${\Delta}E_{\rm vibr}=v_{\rm L}(h/L_X)$ whereas the Heisenberg's
uncertainty for these energies will be
${\delta}E_{\rm vibr}=v_{\rm L}{\delta}p_{\rm vibr}=v_{\rm L}(\hbar/L_X)$, hence the energy step
is always ${\sim}2{\pi}$ times larger than the smearing of each energy separated by this step,
whatever the $v_{\rm L}$ in question.

We note in conclusion that the penetration depth of electromagnetic radiation at the relevant
soft THz frequencies (those shown in Table~\ref{tab:01})
exceeds by far the ``short'' dimensions of the GNBs.

\subsection{Substrate or matrix with gold nanobars for THz source}

\begin{figure}[!t]
\begin{center}
\includegraphics[width=0.45\textwidth]{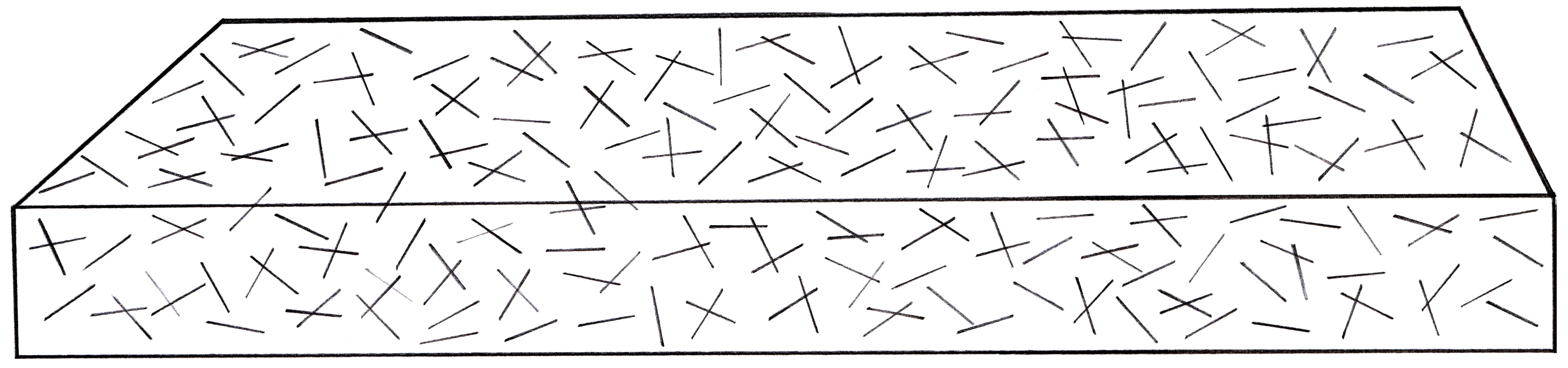}
\end{center}
\caption[example]{\label{fig:04} 
Soft THz radiation source in the form of a matrix transparent in THz 
with randomly aligned embedded GNBs.
}
\end{figure}

For nanobars designed to generate soft THz photons, it is important that they are 
either deposited on a substrate or embedded in a matrix of a material (see Fig.~\ref{fig:04})
that is not only transparent in the soft THz range, but also withstands sufficiently
high temperature heating. Heating nanobars would increase the number of longitudinal 
phonons in them, which would increase the power of the generated THz
radiation. Specifically for the applications in the field of hidden objects detection,
the detection range in the THz domain is limited by the absorption by water vapor,
therefore increasing the THz source power might be an important issue. 

Another way to increase the radiation power is to use a large number of nanobars, 
deposited on / in a large area of substrate (or matrix); this would also facilitate
generating a required large-diameter (${\sim\,}40$~cm) spot.

Heating nanobars could be carried out by microwave radiation, namely, with the help 
of a household microwave oven (see Sec.~\ref{sec:03} below); the dimensions 
of the furnace chamber are large enough to accommodate the substrate / matrix 
of significant size. In addition, the metal walls of the chamber, reflecting
the THz radiation primarily emitted in all directions, would also contribute 
to a more efficient use of the generated radiation. Conditions for heating the GNBs
by microwave radiation see in Appendix B.

Among easily available heat-resistant materials, Teflon\textsuperscript{\textregistered}
is transparent in the THz\cite{THz-Materials}.
The use of a Teflon\textsuperscript{\textregistered} matrix and application 
of a standard 2.45~GHz microwave radiation would allow GNBs to be heated to temperatures
of about 260~$^{\circ}$C. Some matters related to the possible ways of manufacturing 
the Teflon\textsuperscript{\textregistered} matrix were considered in our work.\cite{FANEM2018}
An even greater increase in the temperature of nanobars could be achieved 
by using for the substrate or matrix such materials as the high-resistivity float-zone silicon,
crystal quartz, or sapphire.\cite{THz-Materials}

\section{Possible design of the gold-nanobars-based soft TH\lowercase{z} radiation source}
\label{sec:03}

\begin{figure}[ht]
\begin{center}
\includegraphics[width=0.40\textwidth]{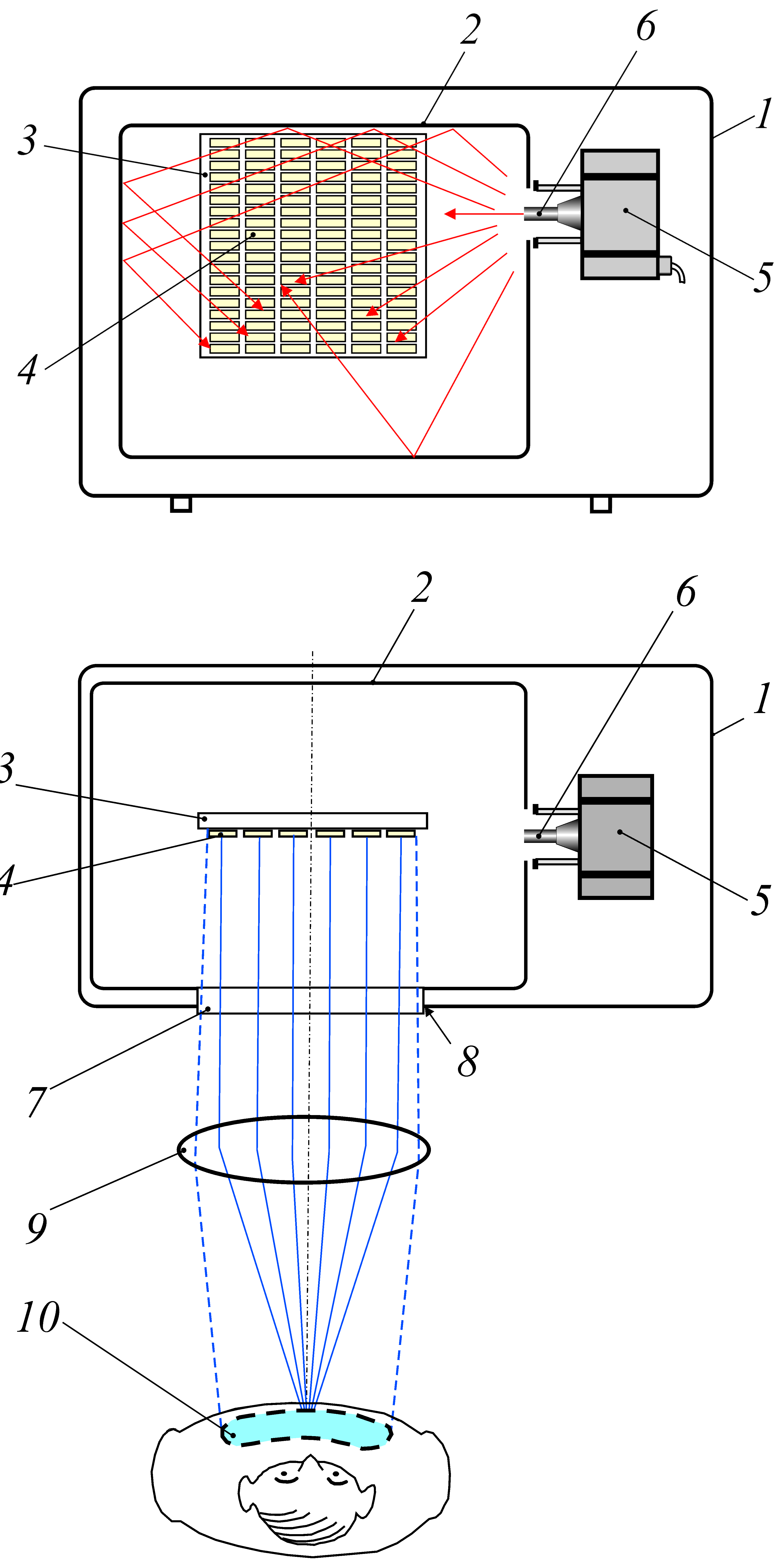}
\end{center}
\caption[]{\label{fig:05} 
Possible construction of the GNBs-based soft THz radiation source.
Upper panel: front view, lower panel: top view. See text for discussion.}
\end{figure}

The soft THz radiation source could be designed as shown in Fig.~\ref{fig:05}
(see also patents Ref.~\citenum{Patent-RU2622093,Patent-RU2650343}).
The housing 1 with metal chamber 2 inside comprises a substrate 3 
with GNBs 4 deposited on it (otherwise, a matrix can be used with embedded GNBs).
An electromagnetic emitter in the form of a magnetron 5 with a waveguide 6
opens into the chamber; the power, control and cooling systems are not shown. 
In the housing 1, an opening 8 is made, in which a resonant filter 7 for the outcoming
THz radiation is installed. 
Beyond the filter 7, a focusing system 9 can be placed, collecting the THz radiation 
emitted by GNBs 4 and focusing it on a hidden object 10 on the subject under examination.
Filters and lenses for the THz range are commercially available and manufactured, for example, 
by the TYDEX\textsuperscript{\textregistered}.\cite{THz-Filters,THz-Lenses}

The source operates as follows. The magnetron generates microwave photons, 
which enter the cavity and heat the GNBs, that is manifested as an increase 
in the number of longitudinal phonons.

\begin{figure}[t!]
\begin{center}
\includegraphics[width=0.48\textwidth]{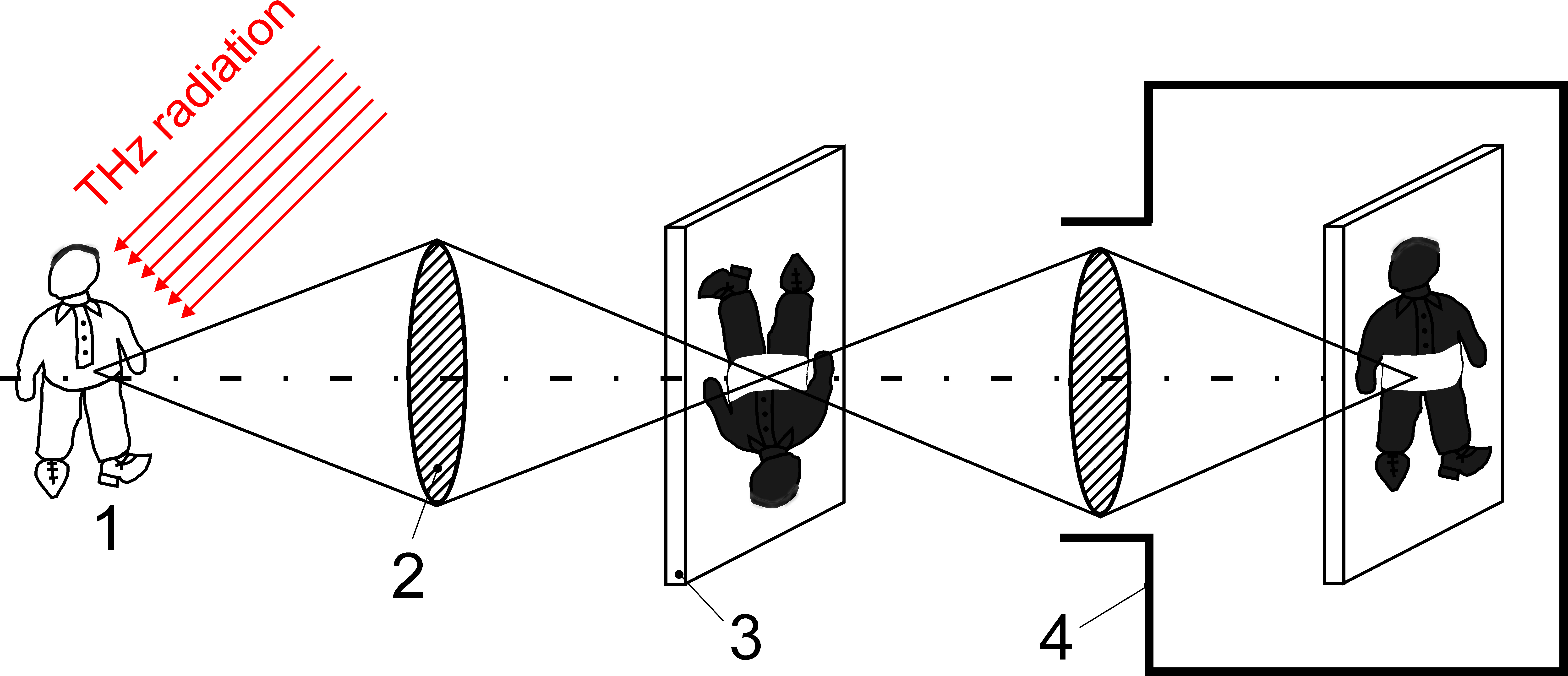}
\end{center}
\caption[]{\label{fig:06} 
Detection and imaging of objects concealed under clothing. 
1: the person inspected; 2: objective focusing the reflected THz radiation 
onto the converter; 3: THz-to-IR converter; 4: IR camera.}
\end{figure}

As was argued above, those Fermi electrons in the GNBs which undergo an absorption / emission of phonons
along the scenario elaborated in Sec.~\ref{sec:02} end up in an excited state 
the relaxation from which can not proceed by exiting the sample (the work function of gold being
prohibitively high) nor via emission of a low-energy (close to the zone center) phonon.

When using a THz radiation source to scan humans, that is, for a non-covert examination
at short distances (${\leq}1$~m) to detect objects hidden under clothing, the focusing system 9 
might be left out. In this case, instead, an unfocused flow of THz radiation is required, 
which would provide a spot with a diameter of ${\sim}40$~cm on the human body.
THz photons reflected from the metal parts of the hidden object are recorded by the THz detector
(see Figs. \ref{fig:01} and \ref{fig:06}). As a possible realization of the latter, 
a THz-to-IR converter could be used, similar to that suggested for visualization of 
malignant tumour.\cite{Ferroel509-158} More detailed estimations of the expected parameters
of this device's performance can be found in Ref.~\citenum{FANEM2018}.

\section{TH\lowercase{z} to IR converter for visualization of hidden objects}
\label{sec:04}
The second essential component of the proposed concealed object detection system is the 
setup for visualization of scattered / reflected THz image; the idea is to convert 
the distribution of intensities over the THz wavefront into the map of temperatures,
and inspect the latter with the help of a standard IR camera. The working element 
of the THz to IR converter is the plate covered with a sufficiently dense array of
gold nanospheres (GNS), to be heated under the effect of incoming THz radiation, 
or the matrix bearing such objects densely embedded within.

Fig.~\ref{fig:06} outlines the general suggested scheme of the detection setup.
The subject 1 is irradiated with a soft THz radiation source. The radiation reflected 
from the object hidden on the subject's body is focused by the lens 2 on the THz-to-IR converter 3 
and creates the hidden object's image in THz rays on the converter plate. The converter absorbs
THz radiation and channels its energy into excitation of longitudinal phonons, that amounts to
local heating and creation of a point source of IR radiation associated with each 
single gold nanosphere (GNS). The resulting two-dimensional picture of IR ``pixels'' is perceived
by a standard IR camera 4 and made visible on the camera's display. Technical considerations
to be elaborated include $(i)$ the choice of acceptable GNS sizes to ensure a good ``performance''
of a single GNS as heat emitter, in view of reaction time and the spatial dissipation of heat
into the substrate or matrix, that will blur the ``pixel'', and $(ii)$ the distribution of GNSs 
within the converter's plate to ensure sufficient sensitivity and spatial resolution
of thermal picture to be perceived by the IR camera.

\begin{figure}[b!]
\begin{center}
\includegraphics[width=0.48\textwidth]{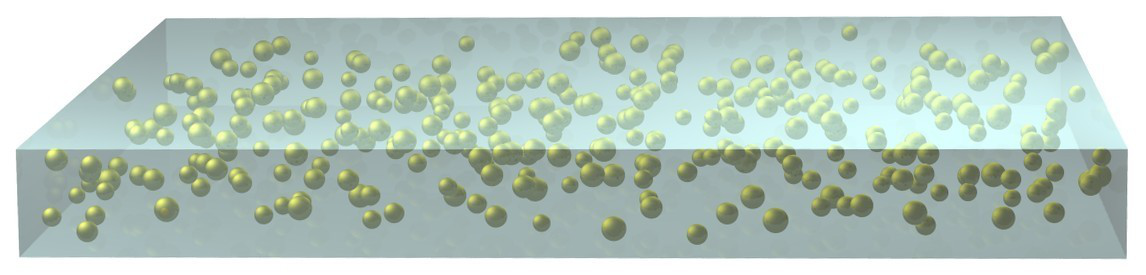}
\end{center}
\caption[]{\label{fig:07} 
THz-to-IR converter in the form of the matrix with embedded GNSs.}
\end{figure}

If the matrix' thickness $\delta$ is within the depth of focus $l_{\rm dof}$ of the IR camera's
objective, the matrix seems preferable over single-layer deposition, because it allows to achieve
larger ``projected'' density of GNSs per surface unit. In the following analysis, 
we assumed $\delta=0.1$~mm (of the Teflon\textsuperscript{\textregistered} film) and 
$l_{\rm dof}\,\approx\,0.3$~mm, therefore, we have done all estimations for the matrix
(Fig.~\ref{fig:07}).

\begin{table*}[t!]
\caption{\label{tab:03}
Parameters of GNSs intended for registration of soft THz radiations,
for the pixel size $d=15$~$\mu$m and the thickness of the Teflon\textsuperscript{\textregistered}
matrix $\delta=0.1$~mm. Concentration of GNSs in the matrix is $(d^2\,\delta)^{-1}=4.44{\cdot}10^4$~mm$^{-3}$.
}
\begin{center}
\begin{tabular}{|p{7.5cm}|c|*{8}{p{0.40cm}|}}
\hline
 & & \multicolumn{8}{c|}{Frequency (THz)} \\
\cline{3-10}
Property & Units &
\multicolumn{2}{c|}{0.14} & \multicolumn{2}{c|}{0.24} & 
\multicolumn{2}{c|}{0.41} & \multicolumn{2}{c|}{0.70} \\
\hline
Phonon momentum & $h/a_{\rm Au}$ &
\multicolumn{2}{c|}{0.020} & \multicolumn{2}{c|}{0.024} &
\multicolumn{2}{c|}{0.050} & \multicolumn{2}{c|}{0.090} \\
Phonon momentum & $10^{-21}$~g$\,$cm/s &
\multicolumn{2}{c|}{3.25} & \multicolumn{2}{c|}{5.35} &
\multicolumn{2}{c|}{8.11} & \multicolumn{2}{c|}{14.60} \\
\hline
\multicolumn{2}{|l|}{$m_{\rm el}$} &
\multicolumn{2}{c|}{36} & \multicolumn{2}{c|}{12} &
\multicolumn{2}{c|}{5} & \multicolumn{2}{c|}{2} \\
\multicolumn{2}{|l|}{$n_{\rm vibr}$} &
\multicolumn{2}{c|}{1} & \multicolumn{2}{c|}{1} &
\multicolumn{2}{c|}{1} & \multicolumn{2}{c|}{1} \\
\hline
$D\,\approx\,4.23(m_{\rm el}/n_{\rm vibr})^{1/2}$ \rule[-2pt]{0pt}{12pt} & nm &
\multicolumn{2}{c|}{25.4} & \multicolumn{2}{c|}{14.65} &
\multicolumn{2}{c|}{9.5} & \multicolumn{2}{c|}{6.0} \\
\hline
Thermal conductivity $\lambda_{\rm 1p}$ \rule[-3pt]{0pt}{13pt} & W$\,$m$^{-1}$K$^{-1}$ &
\multicolumn{2}{c|}{220.1} & \multicolumn{2}{c|}{126.9} &
\multicolumn{2}{c|}{82.3} & \multicolumn{2}{c|}{52.0} \\
\hline
${\Delta}p_D$ & $10^{-21}$~g$\,$cm/s &
\multicolumn{2}{c|}{${\geq}\,0.41$} & \multicolumn{2}{c|}{${\geq}\,0.72$} & 
\multicolumn{2}{c|}{${\geq}\,1.11$} & \multicolumn{2}{c|}{${\geq}\,1.76$} \\
${\Delta}p_{\rm F}$ & $10^{-23}$~g$\,$cm/s &
\multicolumn{2}{c|}{0.66} & \multicolumn{2}{c|}{1.135} & 
\multicolumn{2}{c|}{1.94} & \multicolumn{2}{c|}{3.31} \\
\hline
Transmission \rule[0pt]{0pt}{10pt} of the 0.1~mm thick 
Teflon\textsuperscript{\textregistered} film \rule[-3pt]{0pt}{6pt}
 & \% &
\multicolumn{2}{c|}{$\sim\,$90} & \multicolumn{2}{c|}{$\sim\,$90} & 
\multicolumn{2}{c|}{$\sim\,$92} & \multicolumn{2}{c|}{$\sim\,$92} \\
\hline
\multicolumn{2}{|l|}{Emissivity factor $\alpha$} & 
\multicolumn{1}{c|}{1} & \multicolumn{1}{c|}{0.5} &
\multicolumn{1}{c|}{1} & \multicolumn{1}{c|}{0.5} &
\multicolumn{1}{c|}{1} & \multicolumn{1}{c|}{0.5} &
\multicolumn{1}{c|}{1} & \multicolumn{1}{c|}{0.5} \\
\hline
${\Delta}T_{\alpha}$ & mK & 
\multicolumn{1}{c|}{14} & \multicolumn{1}{c|}{28} &
\multicolumn{1}{c|}{14} & \multicolumn{1}{c|}{28} &
\multicolumn{1}{c|}{14} & \multicolumn{1}{c|}{28} &
\multicolumn{1}{c|}{14} & \multicolumn{1}{c|}{28} \\
\hline
$Q_T$ \rule[-2pt]{0pt}{12pt} & $10^{-10}$~W &
\multicolumn{1}{c|}{5.96} & \multicolumn{1}{c|}{11.91} &
\multicolumn{1}{c|}{3.40} & \multicolumn{1}{c|}{6.79} &
\multicolumn{1}{c|}{2.19} & \multicolumn{1}{c|}{4.38} &
\multicolumn{1}{c|}{1.38} & \multicolumn{1}{c|}{2.75} \\
\hline
\multicolumn{2}{|c|}{\hspace*{0.5mm}\parbox[c]{10.0cm}{
\vspace*{-10pt}
\begin{flushleft}
Threshold number of GNSs \rule[0pt]{0pt}{10pt} within volume element $(d^2\delta)$ \\*[-1pt]
that maps onto a pixel of IR camera \rule[-4pt]{0pt}{6pt}
\end{flushleft}\vspace*{-10pt}
}} &
\multicolumn{2}{c|}{0.08} & \multicolumn{2}{c|}{0.13} &
\multicolumn{2}{c|}{0.21} & \multicolumn{2}{c|}{0.33} \\
\hline
Heating time & $\mu$s & 
\multicolumn{1}{c|}{0.57} & \multicolumn{1}{c|}{0.56} &
\multicolumn{1}{c|}{0.43} & \multicolumn{1}{c|}{0.43} &
\multicolumn{1}{c|}{0.33} & \multicolumn{1}{c|}{0.33} & 
\multicolumn{1}{c|}{0.23} & \multicolumn{1}{c|}{0.23} \\
\hline
Cooling time & $\mu$s & 
\multicolumn{1}{c|}{1.53} & \multicolumn{1}{c|}{1.58} & 
\multicolumn{1}{c|}{1.32} & \multicolumn{1}{c|}{1.39} & 
\multicolumn{1}{c|}{1.38} & \multicolumn{1}{c|}{1.36} & 
\multicolumn{1}{c|}{1.18} & \multicolumn{1}{c|}{1.19} \\
\hline 
\hspace*{-1pt}
\parbox[c]{7.5cm}{\vspace*{-10pt}
\begin{flushleft}
Radius at which the temperature \rule[0pt]{0pt}{9pt} 
in the Teflon\textsuperscript{\textregistered} matrix falls to 
1/10 that in the GNS centre \rule[-3pt]{0pt}{6pt}
\end{flushleft}\vspace*{-10pt}
 } & nm &
\multicolumn{2}{c|}{100.0} & \multicolumn{2}{c|}{62.3} &
\multicolumn{2}{c|}{42.0} & \multicolumn{2}{c|}{27.3} \\
\hline 
\hspace*{-1pt}
\parbox[c]{7.5cm}{\vspace*{-10pt}
\begin{flushleft}
Operating power \rule[0pt]{0pt}{9pt} of the $9.6{\times}7.7{\times}0.1$~mm$^3$ 
Teflon\textsuperscript{\textregistered} matrix with embedded GNSs
of diameter $D$ \rule[-3pt]{0pt}{6pt}
\end{flushleft}\vspace*{-10pt}
} & $\mu$W &
\multicolumn{1}{c|}{195.6} & \multicolumn{1}{c|}{390.9} &
\multicolumn{1}{c|}{111.6} & \multicolumn{1}{c|}{222.9} &
\multicolumn{1}{c|}{ 71.9} & \multicolumn{1}{c|}{143.8} &
\multicolumn{1}{c|}{ 45.3} & \multicolumn{1}{c|}{ 90.3} \\
\hline
\end{tabular}
\end{center}
\end{table*}

\begin{figure*}[!t]
\begin{center}
\includegraphics[width=0.70\textwidth]{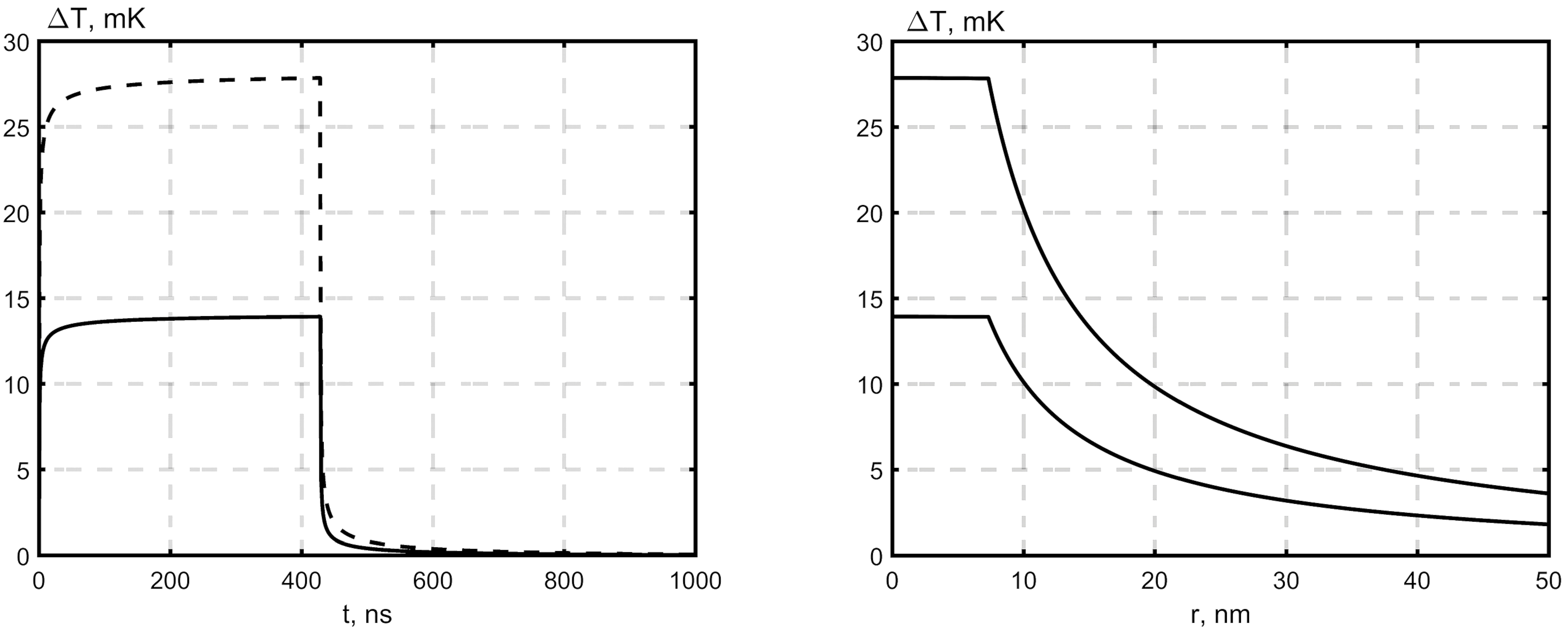}
\end{center}
\caption[]{\label{fig:08} 
Time characteristics (left panel) and radial temperature distributions (right panel) around the GNS 
with a diameter of 14.65~nm (for visualization at 0.24~THz) in the Teflon\textsuperscript{\textregistered}
film matrix for two values of the emissivity factor, $\alpha$=1 (lower curve) 
and $\alpha$=0.5 (upper curve).\bigskip 
}
\end{figure*}

Table~\ref{tab:03} outlines the parameters of GNSs to be embedded into 
the Teflon\textsuperscript{\textregistered} matrix so that to yield the converter's
reasonable efficiency. The basic considerations, including estimations of the GNSs diameters,
are evoked and discussed in Ref.~\citenum{FANEM2018}. In a nutshell, on absorbing a THz photon
by a GNS, the (presumed nearly free) Fermi electron is excited across $m_{\rm el}$ steps of
(confinement-imposed) energy ladder, and then relaxed by releasing a longitudinal phonon with $n_{\rm vibr}$
minimal steps in wavevector and/or energy, assuming the linear dispersion for phonons. 
The mismatch in the momentum conservation on such process, that would normally come about 
due to different dispersion relations for photons, electrons and phonons, will be tolerated 
by force of the Heisenberg's uncertainty relation, as argued in Ref.~\citenum{FANEM2018}.
Specifically, the confinement-conditioned uncertainty of the electron momentum in a particle 
of diameter $D$, ${\Delta}p_D\,\approx\,h/(2{\pi}D)$, should exceed ${\Delta}p_{\rm F}$,
the mismatch of momentum of the Fermi electron on absorption of the THz photon of frequency $\nu$.
The corresponding estimates are listed in Table \ref{tab:03} for four target frequencies 
specified in Sec.~\ref{sec:intro}. For all these frequencies, the soft THz radiation easily penetrates
the particle (cf. ``skin depth'' values in Table~\ref{tab:01}), so that all Fermi electrons 
can participate in the absorption of THz photons.

The upper part of the Table \ref{tab:03} deals with isolated GNSs; the lower part 
concerns the latter's embedding in Teflon\textsuperscript{\textregistered}, as a likely candidate
for the substrate material. The data on the 0.1~mm thick Teflon\textsuperscript{\textregistered}
film transmissions are from Ref.~\citenum{THz-Materials}. The emissivity factor $\alpha$,
a phenomenological property of ``real'' nanoparticles, indicates which part of the energy
delivered will then emerge via IR radiation; the values $\alpha$=1 to 0.5 will presumably bracket 
the realistic estimates. ${\Delta}T_{\alpha}$
is the excess of temperature over the background to be created in the GNS to make it detectable
by the IR camera; the nominal sensitivity value of $\approx\,14$~mK, characterizing modern IR cameras, 
needs to be increased in case of reduced values of $\alpha$. In order to attain such temperature
on the GNS, the power $Q_T$ needs to be delivered, which follows from the solution of the heat
transfer equation. With $Q_T$ as a source term, the heat equation describes the rise of temperature
with time and the radial temperature profile across the GNS and its surrounding medium. 
These properties are depicted in Fig.~\ref{fig:08} for the particle with diameter
$D=14.65$~nm embedded in Teflon\textsuperscript{\textregistered}.

From Fig.~\ref{fig:08} (left panel) and from Table~\ref{tab:03}
one can conclude that THz-to-IR converter built along the concept outlined should possess 
the heating / cooling times acceptable for real-time operation, and operating power in the range of
accessible.\cite{Proc2004FEL-216,NaturePhotonics1-97,PhysPlasmas15-055502,IEEEspectrum-Armstrong}
Thus, the THz sources discussed, as well as the visualization system based on the THz-to-IR converter, could find application in the equipment for detection of hidden objects.

Note that some issues related to improving the efficiency of conversion of THz radiation
into heat by nanoobjects were discussed in our article,\cite{Nanotechnology29-285704} 
and issues related to possible methods of manufacturing a matrix 
from Teflon\textsuperscript{\textregistered} were considered in Ref.~\citenum{FANEM2018}.
Since the matrix with embedded GNSs is not commercially available, we mention a method 
that may be useful in the laboratory conditions. For the matrix, one could take gelatin,
a water soluble protein transparent in the soft THz range. For embedding into the matrix,
one may find it convenient to use commercial GNSs which are sold in the form of water suspensions.

The data on the transparency of the pure gelatin in the THz range are scarce, therefore 
for the estimates one could use the graph of the transmission spectrum of the silver-ion-doped 
gelatin matrix spanning a wavelength range of 0.2~$\mu$m to 1.5~mm from Ref.~\citenum{Nanotech26-121001}.
Figure 4 in that article shows that there is a transmission window in the THz range, which notably
spans the wavelengths corresponding to ``our'' chosen soft THz frequencies (see Table~\ref{tab:01}).
The transmission of gelatin is relatively high (${\sim}\,80\%$) at 1.25~mm wavelength,
but drops down to $\sim\,50\%$ at 0.43 mm. Although at wavelengths larger than 1~mm the transmission
seems to be gradually declining, one can extrapolate the trend to an acceptable value of ${\sim}\,75\%$
at the 2.1~mm wavelength.

\section{Conclusion}
\label{sec:conclu}
Summarizing, we outlined a concept of soft THz radiation sources 
in the form of the matrix (high-resistivity float-zone silicon, crystal quartz, sapphire or Teflon\textsuperscript{\textregistered}), into which the gold nanobars are embedded.
A specificity of soft THz radiation, for which we selected for our analysis the photons 
with frequencies of 0.14, 0.24, 0.41 and 0.70~THz is that it may come about in the course
of two-phonon processes: first, an excitation of the Fermi electrons with the absorption
of the longitudinal phonon, and then the release of a softer longitudinal phonon, whereby the excited electron
retains the difference in the energies of the two phonons -- respectively, 0.58; 0.99; 1.70 and 2.90 meV
for the above frequencies. The relaxation of electrons possessing these excess energy over the Fermi level
occurs via emission of THz photons. To increase the number of longitudinal phonons (and, as a result, 
to increase the emitted THz power), nanoobjects are suggested to be heated with microwave radiation
at a standard (domestic micro oven) frequency of 2.45 GHz. To generate THz photons with 
the frequencies indicated, the gold nanobars with dimensions 5.3~nm$\,\times\,$5.3~nm$\,\times\,$1.318~$\mu$m
seem suitable. Such THz radiation sources being used together with the THz-to-IR converter 
could be, among other possible applications, be used in the equipment for discovering
and visualization of concealed goods.

\appendix

\section{Enhanced surface power density of spontaneous TH\lowercase{z} radiation by gold nanoparticles
due to confinement}
Here we would like to show just a principal possibility of the significant excess
of the Planck limit in the THz range in gold nanoparticles. Namely, the mechanism
of photon emission discussed in Sec.~\ref{sec:02} may be characterized by quite elevated 
surface power density of spontaneous THz radiation. We estimate this effect only by order of magnitude
and within a very simplified model: we assume the nanoparticle to be a sphere with diameter $D=5.3$~nm;
further on, we suppose that the release of a secondary phonon occurs simultaneously with the emission
of a THz photon by the excited electron. 

The $D$ value chosen lets us refer to Fig.~\ref{fig:03} and to the discussion related to quantization
of states due to confinement on this linear size, without entering the details depending on
the exact nanoobject's shape. For numerical estimations, we limit ourselves by the case of 0.24 THz emitted photons.
We estimate the excess, due to spatial confinement, of the radiation emitted over the predictions
of the Planck's law for the black body radiation for two special cases, taking into account 
the uncertainty in the energy of both the longitudinal phonons and the Fermi electrons.

\subsection*{The uncertainty in the energy of longitudinal phonons}
The spatial confinement at the length scale $D$ brings about the uncertainty of the
phonon momentum ${\Delta}p$, ${\Delta}p\,{\cdot}D\,\geq\,\hbar$.
By relation of $p$ to the phonon energy via the (longitudinal) speed of sound
$v_{\rm L}^{\ast}$ within the ``range of interest'' (throughout the FWHM centered at ${\sim}4$~meV,
in our case), the uncertainty of energy is
\begin{equation}
{\Delta}E\,\approx\,v_{\rm L}^{\ast}{\cdot}{\Delta}p \geq \frac{v_{\rm L}^{\ast}\,h}{2{\pi}D}\,.
\end{equation}
For the Fermi electron which absorbed energy $(E_2\!-\!E_1)$, the excited state
will have an uncertainty ${\sim}{\,\pm}({\Delta}E/2)$ around the energy $(E_{\rm F}\!+\!E_2\!-\!E_1)$.
This would amount to this state being characterized by a finite lifetime ${\Delta}t$
such that
\begin{equation}
{\Delta}E{\cdot}{\Delta}t\,{\geq}\,\frac{h}{2\pi}\,.
\end{equation}
Combining the previous relations, the order-of-magnitude estimate of the characteristic
lifetime of the excited electron is
\begin{equation}
{\Delta}t\,{\simeq}\,\frac{h}{2{\pi}{\Delta}E}\,{\simeq}\,\frac{D}{v_{\rm L}^{\ast}}\,,
\end{equation}
that gives a physically meaningful estimate of the ``flight time'' of a phonon across the nanoparticle.
The average emitted THz power can be roughly related to the photon energy $E_2-E_1{\pm}({\Delta}E/2)$
released during ${\Delta}t$:
\begin{eqnarray}
\langle P_{\rm THz}\rangle &=& \frac{E_2-E_1{\pm}({\Delta}E/2)}{{\Delta}t} \nonumber \\
&{\simeq}&
\left[E_2-E_1{\pm}\frac{v_{\rm L}^{\ast}h}{4{\pi}D}\right]\frac{v_{\rm L}^{\ast}}{D}\,.
\end{eqnarray}
In the phononic energy range of FWHM, the speed of sound is
$v_{\rm L}^{\ast}\,{\simeq}\,10^5$~cm/s. The ``energy uncertainty'' term 
${\pm}(v_{\rm L}^{\ast}h)/(4{\pi}D)\,\approx\,1{\cdot}10^{-2}$~meV
can be neglected compared to $E_2-E_1=0.99$~meV.
The emitted power is then ${\langle}P_{\rm THz}{\rangle}\,{\simeq}\,3{\cdot}10^{-11}$~W,
and the surface density of the THz power emitted by the gold nanosphere 
with the diameter $D=5.3$~nm:
\begin{equation}
\frac{\langle P_{\rm THz}\rangle}{4{\pi}(D/2)^2}\,\approx\,3.4{\cdot}10^5\,\mbox{W/m$^2$}\,.
\label{eq:A5}
\end{equation}
In order to make comparison with the prediction by the Planck's theory,
we integrate the function
$$
\epsilon(\lambda)=\frac{2{\pi}c^2}{\lambda^5}\,\frac{h}{\exp\left(\frac{hc}{k{\lambda}T}\right)-1}
$$
over wavelengths throughout the FWHM, i.e., between $\lambda_{\rm min}=6.53{\cdot}10^{-5}$~m
and $\lambda_{\rm max}=7.76{\cdot}10^{-5}$~m, that yields 2.68~W/m$^2$ ($c$ is the speed of light
in vacuum, $h$ the Planck constant, $k$ the Boltzmann constant, $T$ the temperature assumed to be 300~K). 

Note that the estimate Eq.~(\ref{eq:A5})
was performed 
for a single pair of phonons (with an energy difference of 0.99 meV) within the FWHM. 
This turned out already sufficient to demonstrate that the Planck formula
gravely underestimates in the THz range on a nanometric scale, as was already noted
in the works\cite{Nature561-216,Nature567-E12,ACSphotonics5-3082}
devoted to the study of far-field thermal radiation transfer in nanoscale objects with a size smaller
than the Wien's wavelength $\lambda_{\rm W}$ (${\approx}\,10\;\mu$m at $T=300$~K). 
Experimental studies have shown that deviations from the predictions of the Planck's theory can
reach two to seven orders of magnitude.

\subsection*{Accounting for uncertainty in the energy of the Fermi electrons}
Since the Fermi electron velocity in gold $v_{\rm F}=1.4{\cdot}10^8$~cm/s
is three orders of magnitude higher than the speed of sound in the phononic 
energy range of FWHM, $v_{\rm L}^{\ast}\,\simeq\,10^5$~cm/s, a similar argumentation
hints for an excess over the prediction by the Planck's formula by three more orders
of magnitude, yielding, for gold nanosphere with $D=5.3$~nm as above,
${\langle}P_{\rm THz}{\rangle}/[4\pi(D/2)]^2\,{\simeq}\,2.7{\cdot}10^9$~W/m$^2$.
(For comparison: the surface density of power radiated by the Sun's surface is
$\simeq\,7{\cdot}10^7$~W/m$^2$).
Overcoming the Planck limit of the surface power density of radiation inspires attempts 
to generate soft THz radiation using a large number of nanoparticles.
So it would be possible to make a radiation source of sufficiently large total power,
unattainable or difficult to achieve by other ways. By the order of estimate,
the total power of the distributed source radiating into the solid angle $4{\pi}$
at frequency 0.24~THz, with a matrix diameter of 200~mm, a thickness of 1~mm,
with a concentration of $10^5$~mm$^{-3}$ nanoparticles might be $\simeq\,94$~mW.

\section{Choice of optimal size for gold nanobars}
We discuss now the condition for a microwave photon with energy $h{\nu}$
(assuming $\nu$=2.45~GHz, the standard microwave frequency of a domestic oven) 
to be most efficiently absorbed by a gold nano-object with dimensions $L_X=N_x\,a_{\rm Au}$,
$L_Y=N_y\,a_{\rm Au}$, $L_Z=N_z\,a_{\rm Au}$. The absorption excites 
a Fermi electron into a state with the energy
$E_{\rm F}+m_{\rm el}{\Delta}E_{\rm el}$, several ($m_{\rm el}$)
confinement-dependent quantization steps,
${\Delta}E_{\rm el}=\tfrac{4}{3}E_{\rm F}/N$ by force 
of the Kubo formula,\cite{JPSJ17-975,JPhysColloq38-C2-69}
above the Fermi energy $E_{\rm F}$
($N=4N_xN_yN_z$ is the number of univalent gold atoms in the nanoparticle). Following
the excitation, the electron relaxes on releasing a longitudinal phonon, whereby
the nanoparticle is heated. The vibration states are quantized, the smallest
step in the momenta values being that in the direction in which the GNB is the longest,
i.e., $L_Z$. The corresponding step in energy, assuming linear dispersion law
with the longitudinal speed of sound $v_{\rm L}$ defining the slope, will be
${\Delta}E_{\rm vibr}=v_{\rm L}h/L_Z$. The conservation of energy imposes
\begin{equation}
h{\nu} = m_{\rm el}{\Delta}E_{\rm el} = n_{\rm vibr}{\Delta}E_{\rm vibr}\,.
\label{eq:B1}
\end{equation}
for some $m_{\rm el}$ and $n_{\rm vibr}$ integer. The conservation of momentum
cannot be exactly respected, because the dispersion relations for photons
and phonons are markedly different (see Fig.~9.2 and related discussion
in Ref.~\citenum{FANEM2015}). However, the mismatch of momentum can be ``absorbed''
in the uncertainty of the phonon momentum, ${\Delta}p\,{\simeq}\,\hbar/L_Z$,
which comes about as a (yet another) consequence of spatial confinement.
The condition for ${\Delta}p$ to help match the momenta will read
${\Delta}p\,{\geq}\,{\Delta}p_{\rm F}$, whence the condition on $L_Z$
(formulated for  $\nu=2.45$~GHz):
\begin{equation}
L_Z\;{\leq}\;\frac{v_{\rm F}}{2{\pi}\nu}\,\approx 91~\mu\mbox{m}\,.
\end{equation}
Making use of Eq.~(\ref{eq:B1}) formulated in terms of $N_x$, $N_y$, $N_z$, we get
\begin{equation}
m_{\rm el}=\frac{3h{\nu}}{E_{\rm F}}N_xN_yN_z\,,\quad
n_{\rm vibr}=\frac{\nu a_{\rm Au}}{v_{\rm L}}N_z
\end{equation}
and hence
\begin{equation}
\frac{m_{\rm el}}{n_{\rm vibr}}=\frac{3h\,v_{\rm L}N_xN_y}{a_{\rm Au}E_{\rm F}}\,.
\end{equation}
We are interested in the small size of the GNBs, because heating a small GNB
requires less microwave power. This implies the smallness of the numbers $N_x$, $N_y$, $N_z$
as well as of $m_{\rm el}$ and $n_{\rm vibr}$. Assuming for simplicity $N_x=N_y$
and applying the numerical values of $h$, $v_{\rm L}=3.23{\cdot}10^5$~cm/s,\cite{IJAPM3-275}
$a_{\rm Au}=0.408$~nm\cite{AshMerm_book}, $E_{\rm F}=5.53$~eV,\cite{AshMerm_book}
we obtain:
\begin{equation}
\frac{m_{\rm el}}{n_{\rm vibr}}=1.777{\cdot}10^{-2}{\cdot}N_x^2\,.
\end{equation}
We select the minimum value of the $n_{\rm vibr}$ parameter, namely =1, and search for
its compatible small enough integer values of $m_{\rm el}$. It turns out that at $N_x=13$,
$m_{\rm el}=3.003\,\approx\,3$, and for $N_x$=15, $m_{\rm el}=3.998\,\approx\,4$. We retain the smallest
value $N_x$=13 for calculations in the present work; in particular, we split the ${\Gamma}-X$
interval in Fig.~\ref{fig:03}a into 13 intervals. The GNB's width and height are 
$L_X$=$L_Y$=$N_xa_{\rm Au}$=$N_ya_{\rm Au}$=$13{\times}0.408$~nm = 5.3~nm.
From the relations $n_{\rm vibr}{\cdot}{\Delta}E_{\rm vibr}=h{\nu}$ and
${\Delta}E_{\rm vibr}=v_{\rm L}(h/L_z)$, we obtain for the length of GNB: 
$L_z=n_{\rm vibr}\,v_{\rm L}/\nu$, which for the frequency $\nu$=2.45~GHz yields
$L_z$=1.318~$\mu$m ($N_z\,{\approx}\,3230$), whereby ${\Delta}E_{\rm el}\,\approx\,3.38{\cdot}10^{-3}$~meV.

The synthesis of single-crystal gold nanowires with lengths of up to several microns 
has already been mastered.\cite{Nanolett8-2041,AngewChemIntEd49-6156,JACS130-14422,Langmuir24-9855}
Manufacturing the GNBs of 1.318~$\mu$m length will hopefully not be a problem. 
According to the theoretical analysis,\cite{SuperlatNanostruc100-237}
for gold nanowire to be ``stable'', its diameter must be greater than $4.5\,a_{\rm Au}$,
i.e., larger than 1.84~nm. Our estimates of the minimal ``thickness'' of the GNBs 
well respects this condition. 



\end{document}